# Tools and techniques for efficient high-level system design on FPGAs

Adrian J. Chung, Kathryn Cobden, Mark Jervis, Martin Langhammer, Bogdan Pasca
Altera European Technology Centre, UK

*Abstract*—In order for FPGAs to be successful outside traditional markets, tools which enable software programmers to achieve high levels of system performance while abstracting away the FPGA-specific details are needed. DSPB Builder Advanced (DSPBA) is one such tool. DSPBA provides model-based design environment using Matlab's Simulink frontend that decouples the fully-algorithmic design description from the details of FPGA system generation. DSPBA offers several levels of debugging: from Simulink scopes to bit-accurate-simulation and silver reference models. It also offers the most comprehensive set of fixed-point, floating-point and signal-processing IPs available today. The combination of 7 floating-point precisions, fused-datapath support, custom operator support and automated folding allows exploring the best tradeoffs between accuracy, size and throughput. The DSPBA backend protects users from the details of device-dependent operator mapping offering both efficiency and prompt support for new devices and features such as the Arria10 floating-point cores. The collection of features available in DSPBA allows both unexperienced and expert users to efficiently migrate performance-crucial systems to the FPGA architecture.

## I. Introduction

FPGA adoption outside traditional markets requires high levels of system performance while offering turn-around times comparable to competing platforms. Application development – depicted in Figure 1 – has traditionally been divided into three stages: 1/ algorithm development where field application experts together with software engineers prototype the expected behaviour of the system using tools such as Matlab 2/ algorithm implementation where the algorithm developed in the first stage is translated into a platform specific implementation by specialized hardware engineers and 3/ algorithm verification where the implementation correctness is verified against the reference model. Traditional application development maps each stage to a separate group (team), with a fixed schedule of deliverables between the groups. Several iterations between algorithmic specification and implementation may be required for a verified solution meeting performance requirements. Traditionally, implementing an FPGA-based solution using hand-crafted RTL was a complex, tedious and time-consuming task (iterations can take months), justified only by the ultimate performance benefits offer by the platforms. However, this methodology is incapable to offer the shorter time-to-market requirements of less computationally demanding markets that FPGAs could expand into.

DSP Builder Advanced (DSPBA) [1] proposes to break this traditional application development cycle by allowing designers to develop and verify at a high level (steps 1/ and 3/) while obtaining a push-button efficient implementation (step 2/). This significantly shortens the iterations between algorithm design and target-dependent implementation, allowing for early catches of fundamental design issues.

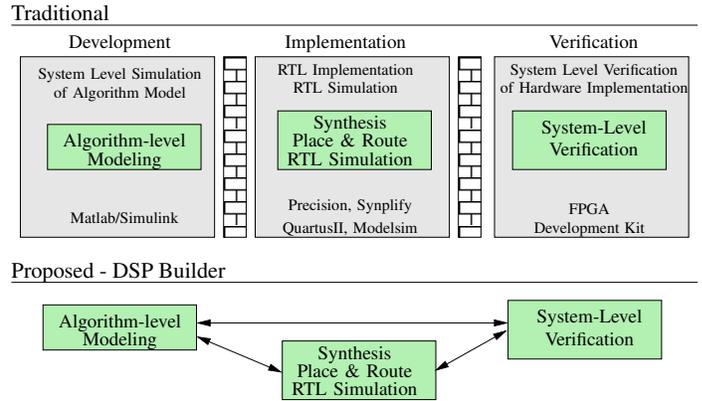

Fig. 1. Traditional System Level Design Flow

DSPBA is high-level design tool with a model-based design entry which integrates with Matlab's Simulink frontend. DSPBA allows users to express their algorithm functionally without having to worry about the FPGA-specific implementation details. The algorithms can be debugged and verified for functionally using classical Simulink scopes and variables, together with DSP Builder-specific simulation features: silver reference and bit-accurate simulations which are described in detail in Section II.

Implementation tradeoffs can be easily explored by means of parametric designs. DSPBA is composed of an extensive set of parametrized library blocks including fixed and floating-point IP together with numerous DSP filters. These blocks, together with the type propagation system build around Simulink, allows the updating of system precision simply by updating input or output precision. The challenges of parametric design are discussed in Section IV.

Further exploration of implementation tradeoffs is possible by means of the fused datapath technology integrated in DSPBA. Entire floating-point datapaths are fused at compile time into a single operator allowing to remove redundant operations and therefore reduce logic and latency. The novel parametrized techniques are presented in Section V.

Application domains such as industrial process data at much slower rates than the FPGA frequency. This potentially allows for hardware reuse during the implementation stage. With DSPBA, users can focus on their algorithm development – not needing to worry about how resources will be shared – and only during the final stages of hardware generation make use of available automatic resource sharing technologies to reduce implementation footprint. The details of the folding solutions are presented in Section VI.

Hardware generation is optimized for the target FPGA and a user-defined frequency. The tool automatically pipelines components and subsystems to match or exceed the target frequency while making target-dependent decisions including the DSP modes that can be used for certain frequencies. The automatic pipelining techniques are presented in Section VII.

The features described in this paper allow users having little or no FPGA knowledge to produce solutions which often outperform hand-crafted RTL while significantly shortening development times. The current paper builds upon [2] which also highlighted a set of high-level design techniques for model-based design; it adds contributions in simulation modelling, parametrized floating-point designs, a new parametrized fused floating-point datapath formulation and next generation automatic pipelining approaches.

## II. SIMULATION MODELLING

### A. Simulink simulation

The design tool is closely integrated with Simulink's simulation system. This allows the library of advanced blocks to be mixed freely with Simulink built-in blocks in the creation of test harnesses, the generation of stimulus data, and the analysis of output signals. A small set of Simulink built-in blocks are also recognized by the synthesis flow when generating HDL and thus can be used anywhere in the design. This allows designs to fully support the aggregate signal types such as complex and vector valued signals. Bus signals allow arbitrary mixtures of signal types on the same wire which provides a convenient mechanism for hiding low level detail.

### B. Multiple Precision and Multiple Representation

DSPBA supports the full range of fixed point types that are provided by Matlab through Simulink toolkit libraries. The design tool also provides an extended set of custom types in order to overcome limitations on precision imposed by the Simulink type system, and to offer the designer more flexibility when managing the trade-off between accuracy and hardware utilization.

The fixed point format is fully parametrized on signed-ness, bit width, and fraction length. The advanced blockset library provides a bit exact simulation over the full range of precisions for all custom blocks. The library is fully orthogonal, allowing arbitrary combinations of fixed point types to be used within the same design. The synthesis flow will also automatically adapt the generated HDL to ensure that timing can still be met as precision is increased, illustrated by the accumulator pipelining in – Figure 2.

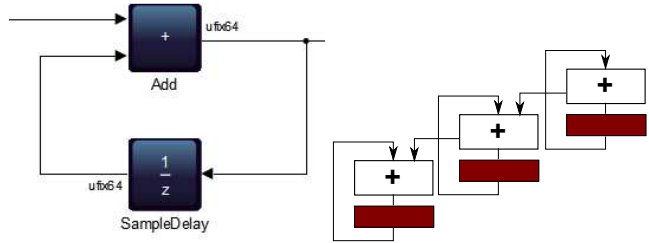

Fig. 2. Single channel accumulator using a wide adder. The design tool automatically splits the adder into adders just wide enough to meet fMAX.

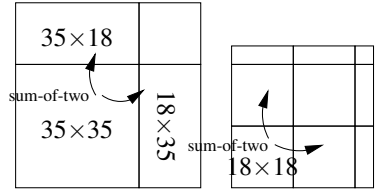

Fig. 3. Wide multipliers are tiled into multiplier precisions that are supported directly by the architecture being targeted. Different tiling patterns are assessed against the multiplier-LUT threshold which controls the tile size that will be mapped to a LUT instead of a DSP block. The clock frequency also affects the availability of certain multiplier precisions, causing the tiling to favor smaller tiles at higher fMAX targets.

### C. Floating point formats

The design tool supports the standard floating point precisions which Simulink provides: IEEE-754 single and IEEE-754 double precision. A subset of blocks in the advanced blockset library can synthesise HDL for processing floating point signals. This includes all the fundamental operations (addition, subtraction, multiplication), as well as a fully featured library of elementary mathematical functions such as trigonometric functions, square root, exponent and logarithm. In addition to the built-in floating point formats, the library extends the range of precisions with 5 extra custom precisions which are presented in Table I.

The library of fundamental and elementary mathematical functions supports every precision, but only a select few blocks (e.g. multiply) will allow different custom precisions to be combined. However, there is a general type conversion block that allows users to mix different precisions in the same design.

We simulate designs that use the built-in floating point formats with the same standard mathematical library that Matlab/Simulink employs. Simulating designs that use the custom precision types however, relies on the special purpose arbitrary precision library MPFR [3] (C library for arbitrary-precision binary floating-point computation with correct rounding). The tool is engineered to synthesize HDL that is fully conforming with the IEEE-754 standard for the fundamental operators, and complies with the OpenCL standard for the elementary mathematical functions. In practice the generated HDL yields an RMS error much better than that specified in the OpenCL standard by a comfortable margin.

TABLE I
SUPPORTED FLOATING-POINT FORMATS IN DSPBUILDER ADVANCED

| Type name | $w_E$, $w_F$ | Range | Smallest | $E_{\text{relative}}$ |
|---|---|---|---|---|
| float16_m10 | 5 , 10 | $(-2^{16}, +2^{16})$ | $2^{-14}$ | $9.77 \cdot 10^{-4}$ |
| float26_m17 | 8 , 17 | $(-2^{128}, +2^{128})$ | $2^{-126}$ | $7.63 \cdot 10^{-6}$ |
| float32_m23 SP | 8 ,23 | $(-2^{128}, +2^{128})$ | $2^{-126}$ | $1.19 \cdot 10^{-7}$ |
| float35_m26 | 8 , 26 | $(-2^{128}, +2^{128})$ | $2^{-126}$ | $1.49 \cdot 10^{-8}$ |
| float46_m35 | 10 , 35 | $(-2^{1024}, +2^{1024})$ | $2^{-1022}$ | $2.91 \cdot 10^{-11}$ |
| float55_m44 | 10 , 44 | $(-2^{1024}, +2^{1024})$ | $2^{-1022}$ | $5.68 \cdot 10^{-14}$ |
| float64_m52 DP | 11 , 52 | $(-2^{2048}, +2^{2048})$ | $2^{-2046}$ | $2.22 \cdot 10^{-16}$ |

The HDL is automatically pipelined according the target clock frequency that is required. Latency across a floating point block adjusts to take the minimum value that will still meet timing. All parallel paths are delay balanced as part of this process.

*D. Accuracy vs Resource utilization tradeoffs*

The user can adjust various configuration options that control trade-offs between accuracy and resource utilization. These can be adjusted on a per-block basis thus allowing a fine grained tuning of accuracy when faced with a strict resource limitation.

The blocks that implement the fundamental floating-point operators can be configured to use either correct rounding or faithful rounding. The RMS error for faithful is about double that of correct rounding but allows for significant reductions in logic utilization. Multiple blocks can be arranged into groups that can be configured to share the same rounding configuration, making it easier for design exploration to be scripted and automated.

Adder trees, and blocks such as the scalar product that generate adder trees, can be configured to generate HDL that sacrifices IEEE compliance in favour of substantially lower latency and register utilization. When enabled, the design tool adopts a strategy that selectively omits normalization stages from adders that gradually exchanges lower accuracy with improved logic utilization in a graceful manner. Users can control how many normalization stages are omitted for each adder tree structure. This technique is especially useful for latency constrained designs and is detailed in Section V.

Users can also convert between any two floating point types anywhere in the design. This allows design exploration where different sections of the computations are carried out at slightly different precisions. The float35_m26 custom type can be used in place of single precision for improved accuracy, or if reduced logic utilization is required, the float26_17 custom type can be used instead.

*E. Design Verification*

For pure fixed-point designs, the user of DSPBA can reasonably expect the behaviour of a Simulink simulation to be identical to the execution of the generated hardware. For floating point designs however, the simulation model at the Simulink level of abstraction will not in general coincide to the hardware simulation. This is especially true if the design makes use of the elementary mathematical library blocks, or the user has configured some of the blocks to exploit the reduced logic of faithful rounding or a fused datapath that is not IEEE compliant. For this reason, the design tool provides a selection of sophisticated verification flows by which the user can assess the accuracy of the generated hardware within a context of a given application domain.

In addition to the 7 floating point precisions, there is also a non-synthesizable custom floating point type of extremely high precision which, for all intents and purposes, can be considered to be practically infinite. Users can use this special data type in the MPFR based simulation of their design to arrive at a close estimate of the gold standard output. As this is technically still an approximation, albeit a highly precise one, it is called a silver reference against which results of the other synthesizable simulations can be compared. The silver reference can be used with the same design that will synthesize to hardware provided it has been made parameterizable on data type. It is not necessary to maintain a separate non-synthesizable model.

*F. Automated test-bench generation*

The design tool will automatically generate a test harness in HDL that will drive the synthesized design from stimulus files and then check the output signals for mismatches. The stimulus files and expected output signals are generated by the Simulink simulation. This is usually sufficient for the verification of pure fixed point designs, however floating point designs require more advanced features, especially when using vector or complex signal types.

DSPBA can import the results of a simulation or hardware run into a MATLAB structure. The import feature preserves the vector and complex groupings of the original signals which gives users the ability to write application specific verification functions with the full flexibility that MATLAB scripts have to offer.

The floating point designs that implement Fast Fourier Transforms (FFT) also benefit from application specific verification scripts. The output takes the form of a frequency response spectrum in which some of the components may approach zero in several places. A RMS (root mean squared) error analysis is not suitable for signal components in isolation, but should be assessed relative to the output power of the entire spectrum.

### III. BIT-ACCURATE SIMULATION

Comparing the silver reference with the imported results of a hardware simulation involves waiting for the two simulations to run to completion. During the early stages in the development of a design, the user will typically use shorter input test data to benefit from shorter turnaround times. Hardware compilation times will begin to dominate this development flow.

For this reason, the design tool offers the user a bit exact simulation mode which runs a simulation on a fixed-point intermediate representation of the original floating-design. This intermediate representation is interpreted and not compiled.

Although this simulation is generally slower, on a cycle by cycle basis, than a hardware simulation that uses fully compiled HDL (e.g. ModelSim), for short runs the design-test-analyze iteration is significantly shortened in practice, since the user doesnt have to wait for a large design to compile. Use of the fixed-point representation also implies that the output will match the hardware simulation exactly bit for bit. The bit accurate simulation mode can be enabled on individual subsystems instead of the entire design, allowing users to select a subset of a larger design on which to focus their development and debugging efforts.

## IV. Parametrized floating-point design

Several factors need to be considered when implementing computational datapaths, including the dynamic range of data, input and requested output accuracy, throughput and available resources. These factors will allow selecting between the datatype(fixed-point, floating-point etc.), precision(single, double, custom etc.) and synthesis style(fully pipelined, iterative, etc.). For example: 1/ fixed-point datatypes are generally used in FPGA designs where data has a low dynamic range whereas floating-point datatypes will be used when the dynamic range of data is high 2/ financial and scientific computation usually require higher than single precisions, 3/ industrial application require lower throughputs from some systems. In this section we assume the floating-point datatype, in a fully pipelined application environment although statements may also hold for other parameter combinations.

The notion of precision (number of bits used to represent information, in floating-point $wF + 1$ bits) is tightly coupled to the notion of accuracy (how close is the current value to its mathematical counterpart). Errors – which cause accuracy to be lost – have several types. Rounding errors occur when an infinitely accurate mathematical result needs to be represented in the given floating-point format. We denote by $p = 1 + wF$ the maximum relative error induced by rounding to nearest is $= 2^{-p-1}$. As rounding errors will build up throughout the algorithm – each basic operation can contribute an equal amount, cancellation can amplify existing errors – a few stages in the calculation the error may be bounded by $Ku$. This which potentially invalidate $i = \lceil \log_2 (Ku) \rceil$ bits of the result. It can be observed here that the final accuracy is influenced by $K$ and $u$ (depends on the precision $p$).

The designer can lower the value of $K$ by balancing the adder trees ( an iterative accumulation will have $i = \zeta(nu)$ – where $n$ is number of terms – whereas a balanced binary tree will have $i = \zeta(\log_2(n)u))$, sorting the terms before the addition – adding from smaller terms first, replacing multiply add pairs with fused multiply-add (for each pair of operations the error introduced will be reduced from $2u$ to $u$), using application specific components [4] which allow storing more precision internally while providing the same interface, etc.

The value of $u$ can be reduced by increasing the precision $p$. Floating-point designs parametrized by precision in DSPBA provide a user knob for adjusting the ratios between accuracy, resource requirements and latency. Our example here is the computing kernel of a QR Decomposition, as part of the a `demo_qrd`[1] demo design in DSP Builder Advanced. The FPGA resources required by the implementation for varying precisions are presented in Figure 4 whereas the accuracy of the implementation is shown in Table II.

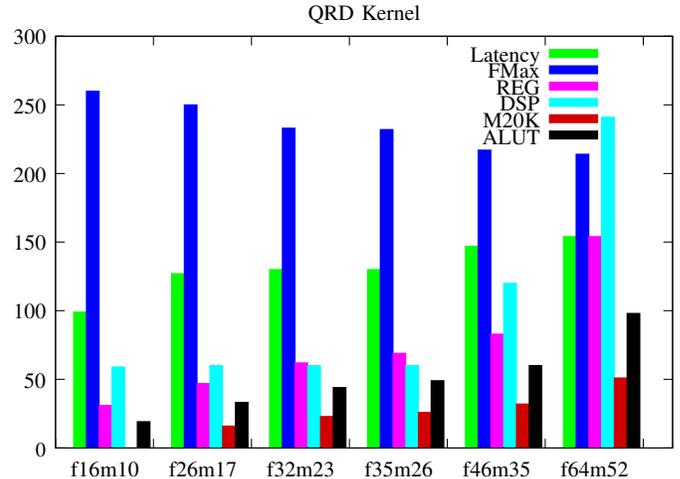

Fig. 4. The resource requirements of the QR Decomposition kernel for varying data precisions. The y axis shows DSPs, M20Ks, Latency, LUTs/$10^3$ and Regs/$10^3$

TABLE II
THE ACCURACY OF QRD WITH VARYING PRECISIONS

| Prec | RMS | Min Rel. | Max Rel |
|---|---|---|---|
| f16m10 | 1.75E-03 | 1.62E-03 | 1.73E+00 |
| f26m17 | 1.33E-05 | 1.41E-05 | 2.31E-02 |
| f32m23 | 2.15E-07 | 1.95E-07 | 1.81E-04 |
| f35m26 | 2.88E-08 | 2.29E-08 | 1.89E-05 |
| f46m35 | 5.34E-11 | 5.33E-11 | 9.03E-08 |
| f64m52 | 1.04E-13 | 1.16E-13 | 2.28E-10 |

The accuracy of the decomposition depends on the size of the input matrix and the data. The decision on which precision to use is therefore governed by the typical or worst case expected data, the available resources for implementation and the required latency of the solution. Assuming that the interface to the rest of the system has a fixed precision, the various precisions can be explored by simply placing convert blocks to and from the the required formats.

For precision exploration to work, all blocks must support precision parametrization. While designing floating-point adders and multipliers is well understood and has been covered in textbooks [5], the design of parametrized functions with architectures targeting contemporary FPGA devices is a much more challenging topic [6]. This is particularly true as the math library in DSP Builder provides full OpenCL coverage [7] – comprised of more than 70 functions – and has passed

---
[1]Demo design can be run by typing demo_qrd in Matlab once the DSPBA libraries have been loaded

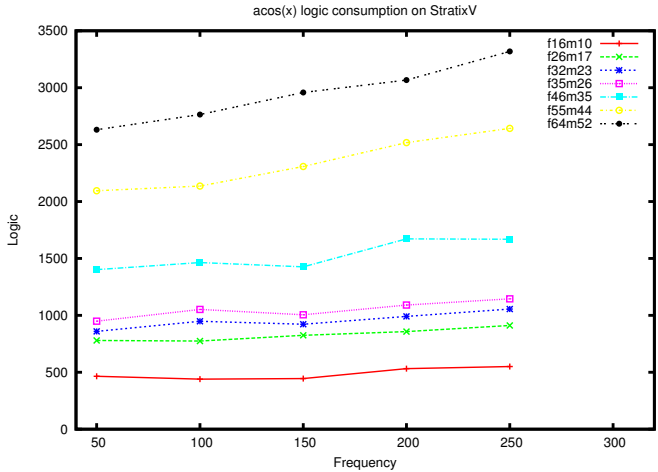

Fig. 5. The logic requirement of a fully parametrized *acos*(x) function across multiple precision and frequency

OpenCL conformance[2].

Large sections of the architecture description of floating-point functions can be easily parametrized. However, most functions will also require the evaluation of a fixed-point function which also needs to be parametrized. The literature provides several approaches for implementing these functions from table-based methods, table-and-multiply methods, CORDIC, quadratic iterations such as Newton-Raphson, Taylor expansions and piecewise polynomial approximations. Table-based methods suffer from bad scalability, CORDIC and Newton-Raphson from lack of generality, and Taylor expansions are not easy to evaluate for all functions. Consequently, the general technique used in DSPBA for implementing these functions is a state-of-the art piecewise polynomial approximation [8] combined with efficient truncated multiplier implementations for Altera FPGAs [9]. The input interval $I$ on which the function in evaluated on is split into a number of subintervals. On each subinterval a polynomial of degree $d$ fixed is used to approximate the function. The evaluation datapath which uses a Horner scheme is shared between all polynomials. The input value is used to detect which polynomial is used by selecting the corresponding coefficients from a ROM. The technique maps perfectly to DSP and embedded memory enabled FPGAs. Its tradeoffs include higher polynomial degrees for fewer subintervals and hence lowering coefficient storage requirements. An example of a parametrized *acos*(x) function is presented in Figure 5.

## V. FUSED DATAPATH DESIGN

Floating-point datapaths can be built by assembling discrete floating-point operators. This emulates the behaviour of microprocessors and GPUs and if only basic operators are used: $(+, -, \times, \div)$ it produces reproducible results between platforms. However, the cost of assembling these datapaths in the FPGA architecture is high, cost largely due to the floating-point adder architecture. The implementation cost can be reduced if the reproducibility of results is relaxed for basic operators as it is for elementary functions.

In such a context floating-point datapaths be fused. Fusing the datapath operators into a single one allows removing some redundant stages between operations which allows saving resources and latency. Systematic fusing of floating-point datapath has first been introduced by Langhammer in [10]. The techniques presented were FPGA architecture-specific (targeting Stratix-IV 36-bit multipliers) and precision specific (single-precision). The operations that gained most from the fused techniques included adder trees and dot-product operators. Early versions of the tool used these fused datapath techniques and were successfully assessed by BDTI [11].

In DSPBA we use new parametrized fused floating-point datapath techniques. This allows scaling to arbitrary floating-point formats (from half-precision to double-precision) and targeting all FPGA families, from low cost to high performance devices. Additionally, the fused-datapath techniques come in two flavours: 1/a bit-equivalent IEEE-754 version which diverges from the standard by allowing local underflow/overflow (in other words is more accurate), and also comes with selectable rounding modes, and a 2/ a fused implementation which allows saving significantly more resources.

Let $w_E$ and $w_F$ denote the exponent and fraction widths of the working floating-point format ($w_E = 8, w_F = 23$ for single precision). The format used by the datapath compiler includes 3 bits for explicitly encoding the exceptions, $w_E + g$-bits ($g = 2$ typically), biased by $2^{w_E-1} - 1$ for the exponent, and $1 + l + w_F + l$ bits for the fraction, where $l$ depends on the depth of the respective node in the addition tree.

The exponent uses the same biased representation as IEEE-754. It is, however, represented in twos complement format on a wider number of bits, which allows for negative biased exponents (local underflow) or local overflow. The number of guard bits is parametrizable, but is typically set to 2.

The fraction is represented on $1 + l + w_F + l$ bits out of which a) $1 + l$ are the integer bits - the 'hidden' 1 is explicitly stored, and 1/bit per level is added to avoid fraction overflow and b) $w_F + l$ are the fractional bits, with the $l$ trailing bits used to provide a similar accuracy of the results as with using rounding-to-nearest on $w_F$ fraction bits, while omitting the rounding stage in the floating-point adder.

Components such as adders, subtracters or fused addition/subtracters receive inputs on $(w_E + 2, l + 1 + w_F + l)$ bits and output data on $(w_E + 2, (l+1) + 1 + w_F + (l+1))$ bits. Components such as multipliers may receive inputs in $(w_E + 2, 1 + w_F)$ format and produce data in the same format, or may produce outputs in the $(w_E + 2, 1 + 1 + w_F + 1)$ format.

Within the datapath the extended fraction used is not normalized. Removing costly normalization stages from adders/subtracters (and multipliers) saves logic and latency but usually introduces inconsistencies with fully IEEE-754 compliant datapaths. Normalization stages are independant modules, seperate from the other operators, which receive the $(w_E + 2, l + 1 + w_F + l)$ format and output normalized and rounded values in the format ($w_E + 2$, $1 + w_F$).

---

[2]http://newsroom.altera.com/press-releases/nr-altera-sdk-opencl-conformance.htm

TABLE III
COMPARISON BETWEEN 16-ELEMENT DOT PRODUCT IMPLEMENTATIONS USING IEEE-754 OPERATOR ASSEMBLY AND THE PRESENTED FUSED DATAPATH TECHNIQUE ON STRATIX V, TARGETING SINGLE PRECISION AND A CUSTOM 35 BIT FRACTION FORMAT

| Type | Precision | Performance |
|---|---|---|
| IEEE-754 | 8,23 | 59clk@380MHz, 8272ALMs, 16DSPs |
| Fused | | 37clk@449MHz, 4566ALMs, 16DSPs |
| Fused | | 43clk@443MHz, 7123ALMs, 32DSPs |

TABLE IV
NEWTON-RAPHSON $1/\sqrt{x}$. TWO ITERATIONS. TARGET FMAX 200MHZ, SAMPLE RATE 10MHZ

| flat | 103 ALMs, 12 18-bit Mults, 327MHz, 18 cycles |
|---|---|
| folded | 137 ALMs, 2 18-bit Mults, 400MHz, 40 cycles |

TABLE V
SOLAR INVERTER SAMPLE DESIGN AT 16KHZ

| flat | 20324 ALMs, 2 DSPs, 4M10K, 165MHz |
|---|---|
| folded | 2500 ALMs, 1 DSP, 1 M10K, 172MHz |

Cast units may be necessary to and from fully compliant IEEE-754, if interfacing with such systems. Casting from IEEE-754 is trivial and requires decoding the exceptions, widening the exponent by 2 bits and concatenating the leading one to the fraction. Casting to IEEE-754 is only allowed from a normalized ($w_E + 2$, $1 + w_F$) format. It requires encoding the exception in the output IEEE-754 format by checking the bounds of the exponent together with information from the 3 explicit exception wires.

Similarly to the parametrized floating-point function cores described in the previous section, the presented fused datapath kernels are also described using the latency and target insensitive builder layer. This allows for building fused datapaths for arbitrary precisions which make also use the best LUT, RAM and DSP modes and which can be pipelined to arbitrary user-defined frequencies.

The performance of the fused datapath techniques is depicted on a dot-product example in Table III. The table depicts the results for a 16-element dot product targeting Stratix V with two floating-point formats: single-precision and a custom precision with 10 exponent bits and 35 fraction bits. As expected, the resource and latency savings are significant when comparing to IEEE-754 datapaths. The presented technique is therefore an efficient method for reducing latency and resource consumption in contexts where result reproducibility is not a requirement.

## VI. RESOURCE SHARING

Resource sharing is a general technique for reducing the footprint of designs where the data rate is slower than the clock rate. Manual design of systems sharing resources can lead to very efficient implementations [12] for one point in the data/clock rate space, but: 1/ lack flexibility – requirements often change from functional to data rate, and 2/ require significant knowledge to design. Automatic resource sharing approaches allow users to design and debug at functional level then input synthesis requirements at RTL generation time.

Two different approaches to generating the final architecture are possible, depending on the data/clock rate factor $f$: when $f < 10$ time-division multiplexing techniques (TDM) are required, whereas when $f > 100$ custom processor architectures are more efficient. Folding factors in the range $10 < f < 100$ require fused approaches, where the computational units are no longer scalar primitives but are entire design subsection being reused.

Resource sharing based on TDM is used when the clock rate is an integer multiple of the clock rate. Shared units operate at clock rate on inputs which are serialized while outputs are correspondingly deserialized. Choosing the most efficient implementation is challenging as the cost of the serializing/deserializing multiplexers sometimes outweighs the cost of the unit being reused while increasing latency. In order to find an efficient solution a simulated annealing algorithm searching for the minimum cost solution is injected with component utilization costs. The results presented in Table IV show the efficiency of the design on a 2-Newton-Raphson iteration fixed-point inverse square root design. The TDM resource sharing is available using the *resource sharing folder* in DSPBA.

For designs where $f > 100$ a custom processor based approach is more efficient. The computational units of the custom processor are extracted from the original design. During the mapping stage 1/ components (such as multipliers) operating on different data types are remapped to the same computational units 2/ logical components having low implementation costs are mapped to a single, fused logical unit 3/ floating-point operations such as n-input adders are split into trees of binary adders etc. A schedule is created using execution information from the original design and the latency of the functional units. Finally, a program which controls the execution of the distributed control unit is generated. The custom processor resource sharing in available in DSPBA using the *ALU Folder* block which inputs the folding factor $f$. Table V reports the performance of the ALU Folder on sample solar inverter designs used in the industrial space.

## VII. AUTOMATIC REGISTER INSERTION AND RETIMING

DSPBA offers the user the ability to choose a particular FMax, and the tool will automatically insert registers in order to meet (or exceed) that target, whilst scheduling operations to ensure functional correctness, and aiming at minimizing implementation footprint.

Automatic pipelining is a constantly evolving feature; the current approach uses separate passes for register insertion (using a greedy algorithm) and scheduling, while a combined approach currently under development – and showing promising results– inserts registers and schedules the operations in one integrated pass.

## A. Greedy approach

The FMax-sensitive automatic pipelined mechanism in DSPBA is similar to that described in [13], and broadly involves walking the datapath, accumulating fractions of a clock cycle of latency (known as sub-cycle latency), and inserting a register at each point where the accumulated clock cycle exceeds the clock period. The approach is greedy, in that it does not consider better places to insert the register; these include narrower wires, or places where the register will be absorbed. In adapting this DAG based method the main problems requiring attention are: the presence of our sample delay blocks, designs with feedback loops in the datapath, and constructing a sub-cycle latency model for every block in our blockset.

Sample delay blocks force the constraint that signals beyond them must arrive a number of cycles later than they are; for example, an adder with a sample delay of -1 on the second input indicates that data sample 1 at time t should be added to data sample 2 from time t-1. The scheduler will re-distribute these delays in a way to try and minimise the occupied area on the chip; so it is not certain that a sample delay will exist in the same place, or indeed of the same amount, in the synthesized design. As such, when accumulating delay we cannot simply assume that the sample delay will be a register/memory – instead, the pessimistic approach assumes the delay is not there in any way and the scheduler stage will ensure functional correctness.

Feedback loops in the datapath cause a problem as the greedy algorithm we use will infinitely accumulate latency around a loop if the loop is not broken. This leads to the problem of where best to break feedback loops in order for the sub-cycle latency of each block to be counted only once. This in itself has its own field of research, so we use the approach where loops are broken as they are detected.

Finally, constructing the sub-cycle latency model for each block takes time as it needed to be done empirically and by understanding how the blocks map to hardware. The biggest problem here is how to model the routing delay between elements on the FPGA – in particular, DSP and memory blocks living in fixed columns generally take longer to route to/from.

The greedy sub-cycle register insertion combined with the ILP scheduler provides a more sensitive latency response especially for lower frequency designs. The comparison against a threshold-based register insertion strategy is depicted for an entire library of operators in Figure 6. As expected, a significant amount of time was spent for tuning the algorithm to perform better in situations where the greedy approach falls short, and where cycles were broken at incorrect locations.

## B. Combined approach

Our combined approach is based on the register retiming algorithm introduced by Leiserson and Saxe [14]. We formulate the design as a timing graph, with specific rules for each block as to how many variable nodes it is represented by (generally, this is one, but can be different as some of our blocks have fixed amounts of latency), and representing

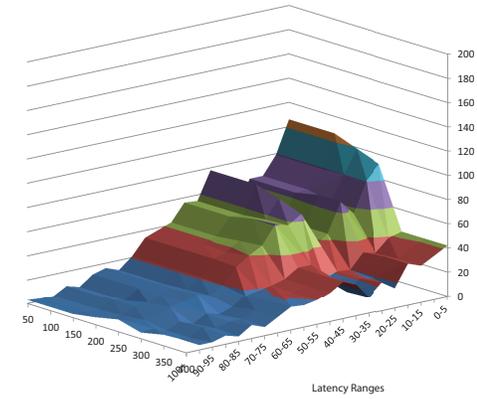

(a) Threshold register insertion + scheduler

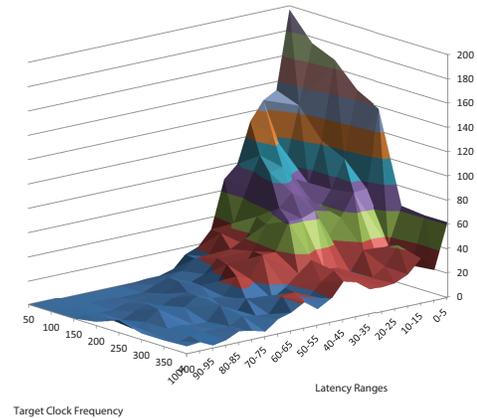

(b) Subcycle register insertion + scheduler

Fig. 6. Comparison between the threshold based scheduler and a subcycle based scheduler on a comprehensive set of designs. For the same fMax the latency is much reduced (the low latency bins have more elements) and the fidelity of the tuning is greater. The long horizontal section from a) have gone into (b)

sample delays as weights on edges. We then generate a set of constraints on data dependencies between these variable nodes, and a set of constraints on the clock constraints between nodes. Generating the clock constraints is computationally complex since we must enumerate every pair of nodes in the graph for which there exists a path between them which has a combined sub-cycle latency of over one clock cycle, so we must be careful which algorithm we use to do this so we dont notice an impact on run-time. Once we have collected both types of constraints we can formulate the entire problem as an ILP problem, consisting only of constraints of the form $a - b \geq c$ (known as difference constraints), as these can be solved more efficiently than multi-variate ILP constraints. We are still evaluating the combined approach but initial results are promising; we should insert registers in more efficient places which should lead to lower resource usage, and, potentially, higher FMax.

## VIII. CONCLUSION

In this paper we have presented a set of features present in the DSPBA tool which allows users with little or no FPGA knowledge to design efficient application systems. Multiple customer designs have shown that 100X to 1000X productivity gains can be expected, compared to a traditional RTL approach. Differing implementation approaches can be quickly iterated using tools level directives. To meet a user defined performance target, the DSPBA tool will automatically adjust the structure of both low level components, such as the pipelining of accumulators, and system level structures, such as the retiming of loops. Mixed precision fixed and floating point datapaths can be specified, and the floating point formats can be further adjusted for accuracy or area optimization over the standard IEEE754 formats. In addition, the fused datapath methodolody used by the tool combines large numbers of discrete floating point operators into a single, highly efficient structure by removing inter-operator arithmetic redundancies. The usable asbtraction of the design process, combined with the effective and efficient implementation results, allows non-traditional FPGA users to target non-traditional FPGA applications.

## IX. ACKNOWLEDGEMENTS

We would like to thank Steven Perry and all engineers who have or are still contributing to this project.

## REFERENCES

[1] "DSP Builder – Advanced blockset with timing-driven Simulink synthesis," 2011, http://www.altera.com/products/software/products/dsp/adsp-builder.html.
[2] S. Perry, "Model based design needs high level synthesis: a collection of high level synthesis techniques to improve productivity and quality of results for model based electronic design," in *Conference on Design, Automation and Test in Europe*, ser. DATE '09, 2009, pp. 1202–1207. [Online]. Available: http://portal.acm.org/citation.cfm?id=1874620.1874909
[3] "MPFR library: multiple-precision floating-point computations with correct rounding," http://www.mpfr.org/.
[4] F. de Dinechin, B. Pasca, O. Creţ, and R. Tudoran, "An FPGA-specific approach to floating-point accumulation and sum-of-products," in *IEEE International Conference on Field-Programmable Technology*. IEEE, 2008, pp. 33–40.
[5] M. D. Ercegovac and T. Lang, *Digital Arithmetic*. Morgan Kaufmann Publishers, 2004.
[6] B. Pasca, "High-performance floating-point computing on reconfigurable circuits," Ph.D. dissertation, École Normale Supérieure de Lyon, Lyon, France, Sep. 2011. [Online]. Available: http://tel.archives-ouvertes.fr/docs/00/65/41/21/PDF/Bogdan_PASCA-Calcul_flottant_haute_performance_sur_circuits_reconfigurables_2011.pdf
[7] Khronos OpenCL Working Group, *The OpenCL Specification, version 1.0.29*, 8 December 2008. [Online]. Available: http://khronos.org/registry/cl/specs/opencl-1.0.29.pdf
[8] F. de Dinechin, M. Joldes, and B. Pasca, "Automatic generation of polynomial-based hardware architectures for function evaluation," in *International Conference on Application-specific Systems, Architectures and Processors*. France Rennes: IEEE, Jul 2010.
[9] S. Banescu, F. de Dinechin, B. Pasca, and R. Tudoran, "Multipliers for floating-point double precision and beyond on FPGAs," *SIGARCH Comput. Archit. News*, vol. 38, no. 4, pp. 73–79, Jan. 2011. [Online]. Available: http://doi.acm.org/10.1145/1926367.1926380
[10] M. Langhammer, "Floating point datapath synthesis for FPGAs," in *Field Programmable Logic and Applications*, sept. 2008, pp. 355 –360.
[11] I. Berkeley Design Technology, "An Independent Analysis of Altera's FPGA Floating-point DSP Design Flow," 2011.
[12] B. Pasca, "Low-cost multiplier-based FPU for embedded processing on FPGA," in *International Conference on Field Programmable Logic and Applications*. Munich Germany: IEEE, 09 2014.
[13] F. de Dinechin and B. Pasca, "Designing custom arithmetic data paths with FloPoCo," *IEEE Design and Test*, 2011.
[14] C. Leiserson and J. Saxe, "Retiming synchronous circuitry," *Algorithmica*, vol. 6, no. 1-6, pp. 5–35, 1991. [Online]. Available: http://dx.doi.org/10.1007/BF01759032